\documentclass[11pt,prd,aps,amssymb,amsmath,tightenlines,showpacs,twocolumn]{revtex4}
\usepackage{graphicx}

\usepackage[T1]{fontenc}
\usepackage[latin1]{inputenc}
\usepackage[english]{babel}
\usepackage{amsmath}
\usepackage{amsfonts}
\usepackage{amssymb}
\usepackage{dsfont}
\usepackage{color}

\def\cP{\mathcal P}
\def\cT{\mathcal T}
\def\cC{\mathcal C}
\def\cPT{\mathcal{PT}}
\def\cCPT{\mathcal{CPT}}
\def\one{\mathds 1}

\begin{document}
\title{Two- and four-dimensional representations of the $\cPT$- and
$\cCPT$-symmetric fermionic algebras}

\author{Alireza Beygi$^1$}\email{beygi@thphys.uni-heidelberg.de}
\author{S. P. Klevansky$^1$}\email{spk@physik.uni-heidelberg.de}
\author{Carl M. Bender$^2$}\email{cmb@wustl.edu}

\affiliation{$^1$Institut f\"{u}r Theoretische Physik, Universit\"{a}t
Heidelberg, Philosophenweg 12, 69120 Heidelberg, Germany\\
$^2$Department of Physics, Washington University, St. Louis,
Missouri 63130, USA}

\begin{abstract}
Fermionic systems differ from their bosonic counterparts, the main difference
with regard to symmetry considerations being that $\cT^2=-1$ for fermionic
systems. In $\cPT$-symmetric quantum mechanics an operator has both $\cPT$ and
$\cCPT$ adjoints. Fermionic operators $\eta$, which are quadratically nilpotent
($\eta^2=0$), and algebras with $\cPT$ and $\cCPT$ adjoints can be constructed.
These algebras obey different anticommutation relations: $\eta\eta^\cPT+
\eta^\cPT\eta=-\one$, where $\eta^\cPT$ is the $\cPT$ adjoint of $\eta$, and
$\eta\eta^\cCPT+\eta^\cCPT\eta=\one$, where $\eta^\cCPT$ is the $\cCPT$ adjoint
of $\eta$. This paper presents matrix representations for the operator $\eta$
and its $\cPT$ and $\cCPT$ adjoints in two and four dimensions. A
$\cPT$-symmetric second-quantized Hamiltonian modeled on quantum electrodynamics
that describes a system of interacting fermions and bosons is constructed within
this framework and is solved exactly.
\end{abstract}
\maketitle

\section{Introduction}\label{s1}
A complex Hamiltonian that is $\cPT$ symmetric (invariant under space-time
reflection) may exhibit two phases separated by a phase-transition point, an
unbroken-$\cPT$-symmetric phase in which the energy spectrum is entirely real
and a broken-$\cPT$-symmetric phase in which the spectrum is partly real and
partly complex \cite{R1}. Complex Hamiltonians have been studied extensively
in quantum mechanics and in quantum field theory. Most of this work has
been devoted to the study of {\it bosonic} theories, for which $\cT^2=1$.
However, $\cT^2=-1$ for fermionic theories, a crucial feature that leads to
differences in the formalism. For example, if the Hamiltonian $H$ has a real
eigenvalue, then $H$ has a corresponding degenerate pair of eigenvectors, $\phi$
and $\cPT\phi$; this is a consequence of Kramer's theorem for ordinary quantum
mechanics. Non-Hermitian fermionic systems have been studied within the wider
framework of pseudo-Hermiticity \cite{R2}.

A previous paper \cite{R3} investigated a matrix representation of a nilpotent
fermionic operator $\eta$ satisfying $\eta^2=0$ together with an adjoint
nilpotent operator, denoted generically by $\bar\eta$. These operators satisfied
a fermionic anticommutator relation $\eta\bar\eta+\bar\eta\eta=\epsilon\one$.
The value $\epsilon=0$ corresponds to a Grassmann algebra and the value
$\epsilon=1$ corresponds to a standard fermionic operator anticommutation
relation. However, the value $\epsilon=-1$ was obtained for this anticommutation
relation in a specific case of a four-dimensional matrix. Subsequently, Cherbal
and Trifonov formalized this result \cite{R4}, making use of the non-Hermitian
formulation of quantum mechanics in Ref.~\cite{R2} and the notation of
Ref.~\cite{R5}.

The problem with determining the value of $\epsilon$ for the anticommutator lies
in the definition of the adjoint nilpotent element $\bar\eta$. In Ref.~\cite{R3}
$\bar\eta$ was chosen to be the $\cPT$ {\it reflection} of $\eta$; that is,
$$\bar\eta=\cPT\eta\cT^{-1}\cP^{-1}.$$
This paper revises the definition of $\bar\eta$ in order to make it consistent
with the concept of a fermionic inner product. With this revision, the fermionic
algebra using $\eta^\cPT$, the $\cPT$ adjoint of $\eta$, always gives rise to an
anticommutation relation with $\epsilon=-1$. However, if we use $\eta^\cCPT$,
the $\cCPT$ adjoint, the fermionic algebra becomes the conventional Hermitian
fermionic algebra $\epsilon=1$.

Knowing the structural properties of the fermionic operators is a
technical but important issue as it provides the basis for constructing theories
of many-body systems in second quantization. It is particularly useful in the
context of a given symmetry, such as $\cPT$ symmetry, because the Hamilton or
Lagrange functions constructed in this way automatically have the symmetry 
properties required. The second-quantized approach enables one to describe and
analyse dynamic systems. We illustrate this formalism with an exactly solvable
model of a $\cPT$-symmetric Hamiltonian for fermions interacting with bosons.
This model is based on the structure of quantum electrodynamics. We solve this
Hamiltonian exactly for the eigenvalues and calculate the renormalized mass of
the fermion.

This paper is organized as follows. In Sec.~\ref{s2} we review the choice of the
inner product in order to set our notation and we define the $\cPT$ and $\cCPT$
adjoints using appropriate definitions of the $\cPT$ and $\cCPT$ inner products.
In Secs.~\ref{s3} and \ref{s4}, we investigate two- and four-dimensional
operator algebras and seek a general {\it ansatz} for the matrix representation
of $\eta$ and its respective $\cPT$ and $\cCPT$ adjoints, which we denote as
$\eta^\cPT$ and $\eta^\cCPT$. In Sec.~\ref{s5} we present our
calculation of a simple model of a $\cPT$-symmetric Hamiltonian of fermions
interacting with bosons. Concluding remarks are made in Sec.~\ref{s6}.

\section{$\cPT$ and $\cCPT$ adjoints of fermionic operators}\label{s2}
In this section we follow the approach of Ref.~\cite{R5} and
describe the general (abstract) formulation. 

As the parity operator $\cP$ is linear, its action on the wavefunction of a
finite-dimensional system can be expressed by a matrix $S$, $\cP\psi=S\psi$.
Since parity applied twice yields the identity matrix, it follows that $S^2=1$
and that the matrix $S$ must have the eigenvalues $\pm1$. In contrast, the
time-reversal operator is antilinear, so its action on the system can be
expressed by a matrix $Z$ combined with the complex-conjugate operation on the
function it operates on, $\cT\psi=Z\psi^*$. It is assumed that $[\cP,\cT]=0$.
In terms of these symbols, the $\cPT$ inner product for fermions is defined as
\begin{equation}
(\phi,\psi)_\cPT\equiv(\cPT\phi)^T Z\psi.
\label{e1}
\end{equation}
Thus, the $\cPT$ adjoint of any
operator $A$ is defined by
\begin{equation}
(A^\cPT\phi,\psi)_\cPT\equiv(\phi,A\psi)_\cPT.
\label{e2}
\end{equation}

As was done in Ref.~\cite{R5}, we insert the definition (\ref{e1}) into the left
and right sides of (\ref{e2}), set $A=\eta$, and extract the operator relation
\begin{equation}
\eta^\cPT=S\eta^\dag S.
\label{e3}
\end{equation}
This is the $\cPT$ adjoint for fermionic systems. 

Let us examine the anticommutator of $\eta$ with $\eta^\cPT$. According to
\cite{R2}, we obtain a fermionic algebra with a minus sign:
\begin{equation}
\eta\eta^\cPT+\eta^\cPT\eta=-\one.
\label{e4}
\end{equation}
The minus sign is a signal that the $\cPT$ inner product is not positive
definite.

Following \cite{R1}, one needs to introduce an additional
operator $\cC$ in order to change the $-$ sign in (\ref{e4}) to a $+$ sign.
This operator thus reflects the sign of the norm
\footnote{The mathematical properties of the $\cC$ operator
resemble those of the charge-conjugation operator of Dirac, but in this context
$\cC$ plays a completely different role, simply forcing the norm of the state
vectors to be positive.}. The operator $\cC$ is linear; it is
thus represented by a matrix $K$. Then the
$\cCPT$ inner product is defined as \cite{R5}
$$(\phi,\psi)_\cCPT=(\cCPT\phi)^T Z\psi=(KSZ\phi^*)^T Z\psi\nonumber$$
and, after some algebra, this takes the form
$$(\phi,\psi)_{\cCPT}=\phi^\dag SK\psi.$$
As a consequence, $A^\cCPT$, the $\cCPT$ adjoint of an operator $A$, is defined
by
$$(A^\cCPT\phi,\psi)_\cCPT=(\phi,A\psi)_\cCPT,$$
and thus $A^\cCPT$ is given by the operator relation
$$A^\cCPT=KS A^\dag SK.$$
The $\cCPT$ adjoint is related to the $\cPT$ adjoint by
$$A^\cCPT=KA^\cPT K.$$
In accordance with \cite{R2}, the anticommutator of a fermionic operator $\eta$
with its $\cCPT$ adjoint should satisfy a {\it conventional} fermionic algebra
\begin{equation}
\eta\eta^\cCPT+\eta^\cCPT\eta=\one.
\label{e5}
\end{equation}

\section{Two-dimensional $\eta$, $\eta^\cPT$, $\eta^\cCPT$}\label{s3}
\subsection{Real representations of $\eta$ and $\eta^\cPT$}
We seek a two-dimensional matrix representation in which $\eta^\cPT$ is the
$\cPT$ adjoint of $\eta$ in accordance with (\ref{e3}). A general matrix
\begin{equation}
\eta=\left(\begin{array}{cc} a & b\\ c& -a\\ \end{array}\right),
\label{e6}
\end{equation}
whose square vanishes, has a vanishing trace and determinant. Let us assume
that $a, b$ and $c$ are real numbers. The parameter $a$ is fixed by the
determinant condition
\begin{equation}
a^2+bc=0.
\label{e7}
\end{equation}

In two dimensions parity reflection $\cP$ can be represented by $\sigma_x$, a
real symmetric matrix whose square is unity:
$$S=\left(\begin{array}{cc} 0 & 1\\ 1 & 0\\ \end{array}\right).$$
We then find that
\begin{equation}
\eta^\cPT=\left(\begin{array}{cc} -a &b\\ c & a\\ \end{array}\right),
\label{e8}
\end{equation}
which satisfies the nilpotency condition $(\eta^\cPT)^2=0$. Now,
evaluating the anticommutator of $\eta$ with $\eta^\cPT$, we find that
\begin{equation}
\eta\eta^\cPT+\eta^\cPT\eta={\rm diag}(-4a^2).
\label{e9}
\end{equation}
For nonvanishing values of $a$ the anticommutator (\ref{e9}) is negative and
with the choice $a^2=1/4$ it can be normalized to $\eta\eta^\cPT+\eta^\cPT\eta=
-\one$.

Because the right side of (\ref{e9}) is nonpositive the standard fermionic
algebra with $\epsilon=+1$ does not have a $2\times2$ representation. But the
right side of (\ref{e9}) can vanish if we take $a=0$. Thus, the Grassmann
algebra has a nontrivial representation. For example, we may take
$$\eta=\left(\begin{array}{cc} 0 & b\\ 0& 0\\ \end{array}\right).$$
This result differs from the conventional Hermitian fermionic algebra, where the
standard algebra has a representation but the Grassmann algebra does not
\cite{R3}.

\subsection{$\cCPT$ adjoint}\label{11}
We have not specified the Hamiltonian, which is required to calculate the
$\cCPT$ product. Nevertheless, we can still determine the $\cCPT$ transformed
operator $\eta^\cCPT$ that yields the standard fermionic algebra (\ref{e5}). To
do so, we first use the fact that $\cC$ and $\cPT$ commute to obtain a general
form for the matrix $K$
\begin{equation}
K=\left(\begin{array}{cc} g & B\\ A& -g\\ \end{array}\right),
\label{e10}
\end{equation}
where $g,\,A,$ and $B$ are arbitrary real parameters. Since $K^2=\one$, we
obtain the constraint ${g^2+AB=1}$. Hence, the $\cCPT$ adjoint of $\eta$ is

\begin{widetext}
$$\eta^\cCPT=\left(\begin{array}{cc} -ag^2+bgA+cgB+aAB & -2agB+cB^2-bg^2\\
-2agA+bA^2-cg^2& ag^2-bgA-cgB-aAB\\ \end{array}\right).$$
\end{widetext}

The anticommutator of $\eta$ and $\eta^\cCPT$ is then
$$\eta\eta^\cCPT+\eta^\cCPT\eta={\rm diag}(2a^2AB+c^2B^2+b^2A^2).$$
By using the determinant relation (\ref{e7}) we eliminate $a^2$ and find that
${(bA-cB)^2}=1$, which links the parameters $A,B$ to $b,c$. The choice $bA=cB$
seems to yield the Grassmann algebra $\eta\eta^\cCPT+\eta^\cCPT\eta=0$. However,
we shall see in Subsec.~\ref{00} that because $\cC$ and the Hamiltonian commute
the choice $bA=cB$ is ruled out, and we arrive at the same result as in the
conventional Hermitian fermionic algebra.

\subsection{Ground state, excited state, and number operator}
\label{99}
We can normalize the anticommutator $\eta\eta^\cPT+\eta^\cPT\eta$ to $-\one$ by
rescaling $\eta$ and $\eta^\cPT$ by $2a$. In this case (\ref{e6}) and
(\ref{e8}) become
$$\eta=\frac{1}{2a}\left(\begin{array}{cc} a & b\\ c& -a\\ \end{array}\right),
\quad\eta^\cPT=\frac{1}{2a}\left(\begin{array}{cc}-a &b\\ c & a\\
\end{array}\right).$$

We then define the ground state $|0\rangle$ as that state that is annihilated by
$\eta$: $\eta|0\rangle=0$. Using (\ref{e6}), we represent this state as
$$|0\rangle=\begin{pmatrix}-b\\ a\end{pmatrix}.$$
To create the $\cPT$-symmetric state $|1\rangle$ we operate on $|0\rangle$ with
$\eta^\cPT$ and get
$$|1\rangle=\begin{pmatrix}b\\ a\end{pmatrix}.$$

We define the $\cPT$ number operator as
$$N^\cPT=\eta^\cPT\eta$$
and establish by direct calculation that
$$N^\cPT|0\rangle=0,\qquad N^\cPT|1\rangle=-|1\rangle.$$
Evidently, $N^\cPT$ gives the {\it negative} of the state occupation number.
We use this fact in Sec.~\ref{s5} in constructing a second-quantized form of a
$\cPT$-symmetric fermionic Hamiltonian.

\subsection{General two-dimensional $\cPT$-symmetric Hamiltonian}\label{00}
A consistent fermionic $\cPT$ quantum mechanics must satisfy three conditions:
(i) The Hamiltonian must be self-adjoint with respect to the $\cPT$ inner
product for fermions; that is, the definition (\ref{e1}) must hold; (ii) $H$
must commute with $\cPT$; (iii) the $\cPT$ symmetry must be unbroken. The first
two criteria give the following general form for a real Hamiltonian:
\begin{equation}
H=\left(\begin{array}{cc} \alpha & \beta\\ \gamma & \alpha\\ \end{array}\right)
\qquad(\alpha,\,\beta,\,\gamma~{\rm real}).
\label{e11}
\end{equation}
The matrix representations of the parity and time-reversal operators, that is,
$S$ and $Z$ in (\ref{e1}), are given by $\sigma_x$.

In Subsec.~\ref{11} we obtained the matrix representation (\ref{e10}) associated
with the $\cC$ operator. A property of $\cC$ not considered in Subsec.~\ref{11}
is that $\cC$ commutes with $H$. The commutation of $K$ and $H$ forces $g$ to
vanish, so the earlier constraint $g^2+AB=1$ reduces to $AB=1$.

Subsection~\ref{11} concludes that if $bA=cB$, one obtains a representation
for a Grassmann algebra. However, the determinant condition $a^2+bc=0$ implies
that $bc$ is a nonpositive quantity. Therefore, to have $bA=cB$, $AB$ must also
be nonpositive, which contradicts the constraint $AB=1$. Thus, as in the
conventional Hermitian case, the Grassmann algebra does not have a nontrivial
representation.

The eigenvalues of (\ref{e11}) are
\begin{equation}
\lambda_\pm=\alpha\pm\sqrt{\beta\gamma},
\label{e12}
\end{equation}
and the corresponding eigenvectors are
$$|\lambda_+\rangle=\frac{1}{\sqrt{2}}\left(\begin{array}{c}
\sqrt[4]{\frac{\beta}{\gamma}} \\ \sqrt[4]{\frac{\gamma}{\beta}}\end{array}
\right),~~ |\lambda_-\rangle=\frac{1}{\sqrt{2}}\left(\begin{array}{c}
\sqrt[4]{\frac{\beta}{\gamma}}\\ -\sqrt[4]{\frac{\gamma}{\beta}}\end{array}
\right).$$
The formula (\ref{e12}) indicates that if $\beta$ and $\gamma$ are positive, the
symmetry is unbroken; that is, the eigenvalues are real.

It is easy to establish that
$$\langle\lambda_+|\lambda_+\rangle_\cPT=1, \qquad
\langle\lambda_-|\lambda_-\rangle_\cPT=-1,$$
$$\langle\lambda_+|\lambda_-\rangle_\cPT=\langle\lambda_-|\lambda_+\rangle_\cPT
=0.$$
We introduce $\cC$ as a measure of the sign of norm:
$$\cC|\lambda_+\rangle=|\lambda_+\rangle,\qquad
\cC|\lambda_-\rangle=-|\lambda_-\rangle.$$
The matrix representation of $\cC$ is then
$$K=\left(\begin{array}{cc} 0 & \sqrt{\beta/\gamma} \\ \sqrt{\gamma/\beta} & 0
\end{array}\right).$$
For the Hamiltonian (\ref{e11}), the annihilation operator now reads
$$\eta=\frac{1}{2}\left(\begin{array}{cc} 1 & \sqrt{\beta/\gamma} \\
-\sqrt{\gamma/\beta} & -1 \end{array}\right).$$
As expected, $\eta$ is nilpotent and
$$\eta|\lambda_-\rangle=0,\qquad\eta|\lambda_+\rangle=|\lambda_{-}\rangle.$$

We now obtain the $\cPT$ adjoint of $\eta$ as
$$\eta^\cPT=\frac{1}{2}\left(\begin{array}{cc} -1 & \sqrt{\beta/\gamma} \\
-\sqrt{\gamma/\beta} & 1 \end{array} \right).$$
Defining the $\cPT$ number operator to be $N^\cPT=\eta^\cPT\eta$, we
can show that
$$\{N,\eta\}_+=-\eta,\qquad\{N,\eta^\cPT\}_+=-\eta^\cPT.$$
The minus sign implies that the $\cPT$ number operator $N^\cPT$ gives the
negative of the state occupation number, as discussed in Subsec.~\ref{99}.

In addition, we remark that the Hamiltonian of our $\cPT$-symmetric interacting fermions can be
recast as a free bosonic Hamiltonian:
$$ H=  \Delta\lambda (-N^\cPT) + \lambda_-\one, $$ where $\Delta \lambda = \lambda_+ - \lambda_-.$

The anticommutator $\eta\eta^\cPT+\eta^\cPT\eta=-\one$ but if instead we use
the $\cCPT$ adjoint of $\eta$,
$$\eta^\cCPT=\frac{1}{2}\left(\begin{array}{cc} 1 & -\sqrt{\beta/\gamma}\\
\sqrt{\gamma/\beta} & -1 \end{array}\right),$$
we obtain the conventional anticommutator $\eta\eta^\cCPT+\eta^\cCPT\eta=\one$.

\section{Four-dimensional $\eta$, $\eta^{\cPT}$, $\eta^{\cCPT}$}
\label{s4}
A general set of 12-parameter complex nilpotent matrices was proposed in
Ref.~\cite{R3} as
\begin{equation}
\eta=\left(\begin{array}{cccc}
-ch-bg-af & f & g & h\\ -a(ch+bg +af) & af & ag & ah\\
-b(ch+bg+af) & bf & bg & bh\\ -c(ch+bg+af) &cf &cg& ch\end{array}\right),
\label{eq8}
\end{equation}
where $a$, $b$, $c$, $f$, $g$, and $h$ are arbitrary complex numbers.
This form was constructed assuming that the trace of $\eta$ as
well as its determinant must vanish in order to garantee nilpotency. We use the
convention of Ref.~\cite{R5} for the matrix representations of $S$ and $Z$; that
is,
\begin{equation}
S=\left(\begin{array}{cc} I & 0\\ 0 & -I\\ \end{array}\right),\qquad
Z=\left(\begin{array}{cc} e_2 & 0\\ 0 & e_2\\ \end{array}\right),
\label{eq9}
\end{equation}
where $I$ is the $2\times2$ identity matrix, and $e_2$ is
$e_2=\left(\begin{array}{cc} 0 & 1\\ -1 & 0\\ \end{array}\right)$.
The $\cPT$ adjoint of $\eta$ reads
$$\eta^{\cPT}=\left(\begin{array}{cccc} -F^* & -a^*F^* & b^*F^* & c^*F^*\\
f^* & a^*f^* & -b^*f^* & -c^*f^*\\ -g^* & -a^*g^* & b^*g^* &c^*g^*\\
-h^* & -a^*h^* & b^*h^* &c^*h^*\end{array}\right),$$
where $F=ch+bg+af$. As required, $\eta^{\cPT}$ is also
nilpotent. One can evaluate the anticommutator of $\eta$ and $\eta^{\cPT}$. This
is found to be
\begin{widetext}
$$\eta \eta^{\cPT} + \eta^{\cPT} \eta=\left(\begin{array}{cccc} J+|F|^2K & a^*J
-F^*fK& -b^*J-F^*gK &- c^*J-F^*hK\\ aJ-f^*FK & |a|^2J+|f|^2K & -ab^*J+f^*gK &
-ac^*J+f^*hK\\ bJ+g^*FK & ba^*J-g^*fK& -|b|^2J-|g|^2K &-bc^*J-g^*hK\\
cJ+h^*FK & ca^*J-h^*fK & -cb^*J -h^*gK &-|c|^2J-|h|^2K\end{array}\right),$$
\end{widetext}
where $J=|F|^2+|f|^2-|g|^2-|h|^2$ and $K=1+|a|^2-|b|^2-|c|^2$.

To obtain the fermionic algebras, it is necessary that the off-diagonal terms
vanish. This gives the relations
\begin{equation}
a^*F=-f,\quad b^*F=g,\quad c^*F=h.
\label{Gr}
\end{equation}
However, these relations force the diagonal terms to vanish. The
particular choice of $\eta$ in (\ref{eq8}) proposed in Ref.~\cite{R3} is only
suitable for constructing a $\cPT$-symmetric Grassmann
algebra, where the anticommutator $\{\eta,\eta^{\cPT}\}$ vanishes. An example of
an $\eta$ that satisfies the relations (\ref{Gr}) and leads to a Grassmann
algebra is
$$\eta=\left(\begin{array}{cccc} 1 & 1 & i & -i\\ 1 & 1 & i & -i\\
i & i & -1 & 1\\ -i & -i & 1 & -1 \end{array} \right).$$

Let us examine another set of matrices that cannot be obtained from (\ref{eq8}):
\begin{equation}\eta=\left(\begin{array}{cccc}f & 0& \alpha c & \alpha b\\
0 & f & \alpha b^* & -\alpha c^*\\ \beta c^* & \beta b & -f & 0\\
\beta b^*& -\beta c &0 & -f\end{array}\right), 
\label{eq-etaalt}
\end{equation}
where $b$ and $c$ are complex and $f$, $\alpha$, and $\beta$ are real
arbitrary parameters. This ansatz is a block-form construct with
$2\times2$ matrices that ensures that the matrix is traceless in the simplest
possible fashion. In addition, the off-diagonal elements have been chosen to be
scaled Hermitian conjugates of one another, introducing a minimum number of
parameters. Nilpotency of $\eta$ must now be enforced, and leads
to the requirement that
\begin{equation}
f^2+\alpha\beta(|b|^2+|c|^2)=0.
\label{e15}
\end{equation}
Using the matrix representations of $S$ and $Z$ in (\ref{eq9}), we obtain the
$\cPT$ adjoint of $\eta$:
$$\eta^\cPT=\left(\begin{array}{cccc} f &0 & -\beta c & -\beta b\\ 0 & f & -
\beta b^* & \beta c^*\\ -\alpha c^* & -\alpha b & -f & 0\\ -\alpha b^* &\alpha
c &0 & -f\end{array}\right).$$
Equation (\ref{e15}) implies that $\eta^\cPT$ is also nilpotent.

The anticommutator of $\eta$ and $\eta^\cPT$ is
$$\eta\eta^\cPT+\eta^\cPT\eta={\rm diag}\{2f^2-(\alpha^2+\beta^2)(|b|^2+|c|^2)
\},$$ 
and because of (\ref{e15}) this reduces to
$$\eta\eta^\cPT+\eta^\cPT\eta = -{\rm diag}\{(\alpha+\beta)^2(|b|^2+|c|^2)\}.$$
Thus, the anticommutator is nonpositive. The choice, $\alpha=-\beta$, gives rise
to a nontrivial representation for the Grassmann algebra. However, when $\alpha
\neq-\beta$, the above anticommutator with suitable normalization can be written
as
$$\eta\eta^\cPT+\eta^\cPT\eta=-\one.$$

To obtain the standard fermionic algebra we again consider the $\cCPT$ adjoint
of $\eta$ instead of $\eta^\cPT$. We construct the $\cC$ operator as follows. We
note that the commutation of $\cC$ and $\cPT$ gives
\begin{equation}
KSZ=SZK^*,
\label{01}
\end{equation}
where $K$, $S$, and $Z$ are the matrix representations of $\cC$, $\cP$, and
$\cT$. Another characteristic of the $\cC$ operator is that it commutes with
the Hamiltonian. The procedure to construct a general $\cPT$-symmetric
Hamiltonian for fermionic systems is described in Ref.~\cite{R5}. A matrix $K$
that satisfies the two criteria (\ref{01}) and $[\cC,H]=0$ is parametrized as
$$K=\left(\begin{array}{cccc} g & 0 & -\gamma c & -\gamma b\\
0 & g & -\gamma b^* & \gamma c^*\\ \gamma c^* & \gamma b & -g & 0\\
\gamma b^* & -\gamma c & 0 & -g \end{array} \right),$$
where $g$ and $\gamma$ are real numbers.

The requirement $K^2=\one$ leads to the additional constraint
\begin{equation}
g^2-\gamma^2(|b|^2+|c|^2)=1.
\label{e100}
\end{equation}
Having found $K$, we can easily obtain the $\cCPT$ adjoint of $\eta$:
$$\eta^\cCPT = \left( \begin{array}{cccc} D & 0 & -c A & -b A\\
0 & D & -b^* A & c^* A\\ c^* B & b B & -D & 0\\ b^* B & -c B & 0 & -D
\end{array}\right),$$
where
$$D=f g^2+(|b|^2+|c|^2)\gamma(\gamma f+\alpha g-\beta g),$$
$$A=2\gamma fg-\beta g^2+\alpha\gamma^2(|b|^2+|c|^2),$$
$$B=2\gamma fg+\alpha g^2-\beta\gamma^2(|b|^2+|c|^2).$$
Finally, the anticommutation of $\eta$ and $\eta^\cCPT$ reads
\begin{eqnarray}
\eta\eta^\cCPT &+& \eta^\cCPT\eta\nonumber\\
&=& {\rm diag}\{(|b|^2+|c|^2)[2\gamma f+(\alpha-\beta)g]^2\},\nonumber
\end{eqnarray}
where (\ref{e15}) and (\ref{e100}) have been used. Note that the anticommutator
is {\it positive} and with a suitable normalization can be written as
$$\eta\eta^\cCPT+\eta^\cCPT\eta=\one.$$

For completeness, we remark that the ground state can be defined,
as in Subsec.~IIIC, as being the state that is annihilated by $\eta$: $\eta|0
\rangle=0$. Using (\ref{eq-etaalt}), we represent this state as
$$|0\rangle=\begin{pmatrix}f\\ 0 \\\beta c^* \\ \beta b^*\end{pmatrix}.$$
To create the $\cPT$-symmetric state $|1\rangle$ we operate on $|0\rangle$ with
$\eta^\cPT$ and obtain
$$|1\rangle=\begin{pmatrix}\beta(|b|^2+|c|^2)\\ 0\\fc^* \\ fb^* \end{pmatrix}.$$

Following the procedure in Subsec.~IIIC, after normalizing $|0\rangle$ and $|1\rangle$ above, we ascertain by direct calculation that  $N^\cPT|0\rangle=0$ and $ N^\cPT|1\rangle=-|1\rangle$, where we have used
$N^\cPT=\eta^\cPT\eta$ and (\ref{e15}), thus illustrating again that $N^\cPT$ yields the {\it negative} of the state occupation number.

\section{Simple Model Hamiltonian}\label{s5}
\begin{figure}[h!]
\begin{center}
\includegraphics[scale=1.0]{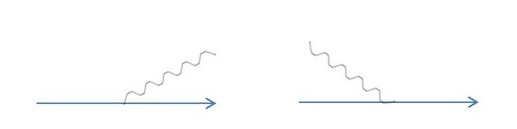}
\end{center}
\caption{The fermion (solid line) can emit or absorb bosons (wavy gray lines).
These are the only possible interactions, so the fermion number is conserved.}
\label{F1}
\end{figure}

In this section we construct a $\cPT$-symmetric model of interacting fermions
and bosons. The idea is based on the Lee model in which the lack of crossing
symmetry makes the model exactly solvable \cite{R6}. We consider a single
fermion that may emit and absorb bosons, as shown in Fig.~\ref{F1}, but the
bosons may not produce a fermion-antifermion pair.

A Hamiltonian that describes this system is
$$H=ma^\dag a-M\eta^\cPT\eta-ga^\dag\eta^\cPT\eta-ga\eta^\cPT\eta,$$
where the operator $a^\dag$ creates (normal) bosons, but the fermionic operator
$\eta^\cPT$ creates a $\cPT$-symmetric fermion. Here, $m$ and $M$ are the bare
boson and fermion masses and g is the coupling amplitude. This Hamiltonian is
not Hermitian but it is $\cPT$ symmetric.

A state containing a single bare fermion and any number $n$ of bare bosons can
be written as $|E\rangle=\sum_{n=0}^\infty c_n|1,n\rangle$. We assume that this
state is normalized; that is,
\begin{equation}
\langle E|E\rangle=\sum_{n=0}^\infty c_n^2<\infty.
\label{e16}
\end{equation}
The annihilation and creation operators for bosons obey $a|n\rangle=\sqrt{n}|n-1
\rangle$ and $a^\dag|n\rangle=\sqrt{n+1}|n+1\rangle$. In addition, $a^\dag
a$ is the boson number operator; that is, $a^\dag a|n\rangle=n|n\rangle$. For
the case of fermions we have the peculiar anticommutation relation $\eta
\eta^\cPT+\eta^\cPT\eta=-\one$. However, as in Sec.~\ref{s3}, we interpret $\eta$
as a lowering operator and $\eta^\cPT$ as a raising operator. Thus, the fermion
number operator is $-\eta^\cPT\eta$.

The time-independent Schr\"odinger equation $H|E\rangle=E|E\rangle$ takes the
form
\begin{eqnarray}
&&\sum_{n=0}^\infty mn c_n|1,n\rangle+\sum_{n=0}^\infty M c_n|1,n\rangle
\nonumber\\ &&\!\!\!\!+\sum_{n=0}^\infty g\sqrt{n+1}c_n|1,n+1\rangle\nonumber\\
&&\!\!\!\!+\sum_{n=0}^\infty g\sqrt{n}c_n|1,n-1\rangle=\sum_{n=0}^\infty Ec_n
|1,n\rangle.\nonumber
\end{eqnarray}
We shift indices and pick off the coefficients of $|1,n\rangle$ to obtain a
recursion relation $c_n$:
$$(mn+M)c_n+g\sqrt{n} c_{n-1}+g\sqrt{n+1}c_{n+1}=Ec_n.$$
The substitution $c_n=d_n\sqrt{n!}$ gives the simpler recursion relation
\begin{equation}
(mn+M)d_n+gd_{n-1}+g(n+1)d_{n+1}=Ed_n.
\label{e17}
\end{equation}

For large $n$ we can neglect the $Md_n$ and $Ed_n$ terms and obtain an
approximate equation for $d_n$ that is valid for large $n$:
$$mnd_n+gd_{n-1}+g(n+1)d_{n+1}\simeq0.$$
There are two consistent asymptotic dominant balances for $n>>1$: If the first
and second terms balance for large $n$, then
\begin{equation}
d_n\simeq(-g/m)^n/n!;
\label{e18}
\end{equation}
if the first and third terms balance, then
\begin{equation}
d_n\simeq(-m/g)^n.
\label{e19}
\end{equation}
(A dominant balance between the second and third terms is inconsistent.)
The norm in (\ref{e16}) becomes $\sum_{n=0}^\infty d_n^2
n!$. Therefore, (\ref{e18}) is acceptable but (\ref{e19}) is not.

Next, we define a generating function $f(x)\equiv\sum_{n=0}^\infty d_n x^n$; if
(\ref{e18}) holds, then $f(x)$ is an entire function of $x$ but if (\ref{e19})
holds, we see that $f(x)$ has a finite radius of convergence with a singularity
in the complex-$x$ plane at $x=-g/m$.

If we multiply (\ref{e17}) by $x^n$ and sum from $0$ to $\infty$, we obtain
the first-order differential equation
$$(mx+g)f'(x)=(E-M-xg)f(x),$$
whose solution is
$$f(x)=Ke^{-gx/m}(mx+g)^{E/m-M/m+g^2/m^2}.$$
As predicted, there is a singularity at $x=-g/m$ unless the exponent in the
second term on the right side is a nonnegative integer $N=0,\,1,\,2,\,\ldots\,$.
This yields the {\it exact} spectrum of physical fermion states:
$$E_N=Nm+M-g^2/m\quad(N=0,\,1,\,2,\,\ldots).$$
Note that as a consequence of the interaction, the mass $M-g^2/m$ of the
physical fermion is {\it lower} than the mass $M$ of the bare fermion.

\section{Brief concluding remarks}\label{s6}
In this paper we have used the alternative formalism for the fermionic scalar
product in Ref.~\cite{R5} to reexamine the operator algebra for fermions in the
context of $\cPT$ symmetry. We have investigated general matrix representations
of the $\cPT$ and $\cCPT$ fermionic creation and destruction operators without
making direct reference to a Hamiltonian. Knowing the behavior of such
operators, especially $\cPT$ operators, can be important for many-body theory,
which often uses the operator definitions to construct the Hamiltonian (in
second-quantized form). It can also be important in understanding the nature of 
species oscillation in neutrinos \cite{R7}.

We have examined the operator algebras in detail for $2\times2$ matrices and for
the $4\times4$ case. Using the algebra that we have developed, we apply the
peculiar anticommutation relations pertinent to the $\cPT$ algebra to construct
a second-quantized $\cPT$-symmetric quantum field theory, namely, a solvable
low-dimensional model of electrodynamics (a modified Lee model) for which the
renormalized energy spectrum is calculated in a closed form and is found to be
real.

\acknowledgments
CMB thanks the Heidelberg Graduate School for Fundamental Physics at Heidelberg University for its hospitality.


\begin{thebibliography}{17}

\bibitem{R1} C.~M.~Bender and S.~Boettcher, Phys.~Rev.~Lett. {\bf 80}, 5243
(1998); C.~M.~Bender, Rep. Prog. Phys. {\bf 70}, 947 (2007).

\bibitem{R2} A.~Mostafazadeh, J.~Phys.~A {\bf 37}, 10193 (2004).

\bibitem{R3} C.~M.~Bender and S.~P.~Klevansky, Phys.~Rev.~A {\bf 84}, 024102
(2011).

\bibitem{R4} O.~Cherbal and D.~A.~Trifonov, Phys.~Rev.~A {\bf 85}, 052123
(2012).

\bibitem{R5} K.~Jones-Smith and H.~Mathur, Phys.~Rev.~A {\bf 82}, 042101 (2010).
See also K.~Jones-Smith, {\it Non-Hermitian Quantum Mechanics} (Dissertation,
Case Western Reserve University, 2010).

\bibitem{R6} T.~D.~Lee, Phys.~Rev.~{\bf 95}, 1329 (1954).

\bibitem{R7} T.~Ohlsson, EPL {\bf 113}, 61001 (2016).
\end{thebibliography}
\end{document}